\documentclass[aps,prc,onecolumn,showpacs,superscriptaddress,preprintnumbers,floatfix,reprint]{revtex4-1}
\usepackage{graphicx}
\usepackage{amsmath}
\usepackage{bm}
\usepackage{yhmath}
\usepackage[colorlinks,citecolor=blue,urlcolor=blue,linkcolor=blue]{hyperref}
\usepackage{subfigure}
\usepackage{color}
\usepackage{cases}
\usepackage{subfigure}
\usepackage{times}
\usepackage{dcolumn,booktabs,bm}
\usepackage{slashed}
\usepackage{amsfonts,amssymb,stmaryrd,latexsym,amsmath}
\usepackage{textcomp}
\usepackage{multirow}
\usepackage{cancel}
\usepackage{array}

\allowdisplaybreaks

\begin{document}

\title{$\Sigma_cN$ interaction in chiral perturbation theory}

\author{Lu Meng}\email{lmeng@pku.edu.cn}
\affiliation{School of Physics and State Key Laboratory of Nuclear
Physics and Technology, Peking University, Beijing 100871, China}

\author{Bo Wang}\email{bo-wang@pku.edu.cn}
\affiliation{Center of High Energy Physics, Peking University,
Beijing 100871, China} \affiliation{School of Physics and State Key
Laboratory of Nuclear
    Physics and Technology, Peking University, Beijing 100871,
    China}

\author{Shi-Lin Zhu}\email{zhusl@pku.edu.cn}
\affiliation{School of Physics and State Key Laboratory of Nuclear
Physics and Technology, Peking University, Beijing 100871,
China}\affiliation{Center of High Energy Physics, Peking University,
Beijing 100871, China}

\begin{abstract}
We adopt the heavy baryon chiral perturbation theory to calculate
the $\Sigma_cN$ interaction to the next-to-leading order. We
consider the contact interactions, one-pion-exchange contributions,
two-pion-exchange diagrams, and renormalization effects of the
vertices, masses and wave functions. With the pion mass dependent
expression, we fit the $\Sigma_cN$ interaction from HAL QCD
calculation with $m_{\pi}\approx 410$ MeV and $m_{\pi}\approx 570$
MeV, and then extrapolate it to the physical pion mass. The
$^3S_1(I=1/2)$ $\Sigma_{c}N$ interaction is weakly attractive but no
bound solution is found. We also propose a quark model to estimate
the leading order $\Sigma_cN$ contact interaction with the $NN$
interaction as input. This approach combining the quark model and
the chiral effective field theory predicts a very attractive
interaction in $^1S_0(I=3/2)$ $\Sigma_{c}N$ channel and a two-body
bound state.
\end{abstract}

\maketitle

\thispagestyle{empty}

\section{Introduction}\label{sec:intro}
In the past decade, many novel hadronic states in the charm sector
have been observed in experiments. Theoretical analysis revealed
that some of these states might be the bound states or resonances of
two heavy hadrons (for recent reviews, see
Refs.~\cite{Chen:2016qju,Esposito:2016noz,Guo:2017jvc,Olsen:2017bmm,Liu:2019zoy,Brambilla:2019esw}).
Thus, the investigation on the hadron-hadron interaction is an
important ingredient to understand these exotic structures.

Nuclear force is the best-understood hadron-hadron interaction,
which has been extensively studied. More than eighty years ago,
Yukawa proposed the one-pion-exchange interaction, which is the
first clear picture about nuclear force~\cite{Yukawa:1935xg}. The
idea was then developed by including other meson-exchange
interactions or operators to obtain the nuclear force with high
precision~\cite{Machleidt:1987hj,Stoks:1994wp,Wiringa:1994wb}. The
modern theory to build the nuclear force is the chiral perturbation
theory (ChPT)~\cite{Epelbaum:2008ga,Machleidt:2011zz}, which is
another inheritor of Yukawa's idea. The formalism originates from
Weinberg~\cite{Weinberg:1990rz,Weinberg:1991um}, in which the
nuclear force can be calculated order by order. The
two-pion-exchange loop diagrams even multi-pion-exchange ones could
be considered by the arrangement of power counting. Apart from these
frameworks at the hadron level, the quark model was also exploited
to calculate the nuclear
force~\cite{Oka:1980ax,Oka:1981ri,Oka:1981rj,Faessler:1982ik,Wang:1992wi,Fernandez:1993hx,Zhang:1994pp,Entem:1999si,Shimizu:1999cr,Nagata:2003gg,Fujiwara:2006yh,Huang:2018rpb}.
Quark model could provide some insights into the origin of the
repulsive core of nuclear
force~\cite{Oka:1981ri,Oka:1981rj,Faessler:1982ik,Shimizu:1999cr}.
Recently, the HAL QCD collaboration proposed a framework to
calculate the nuclear force from the lattice QCD
simulations~\cite{Ishii:2006ec,Aoki:2009ji,Aoki:2012tk}. The methods
mentioned above in calculating the nuclear force were also extended
to study other hadron-hadron
interactions~\cite{Sun:2011uh,Meng:2017fwb,Meng:2017udf,Yang:2018amd,Yang:2019rgw,Chen:2019asm,Guo:2013rpa,Liu:2012vd,Xu:2017tsr,Wang:2018atz,Meng:2019ilv,Wang:2019ato,Ren:2016jna,Song:2018qqm,Haidenbauer:2016vfq,Haidenbauer:2018gvg,Haidenbauer:2019boi}.

Investigating the $Y_cN$ ($Y_c=\Sigma_c$ or $\Lambda_c$) interaction
is a natural extension of the $NN$ interaction, which is important
to explore the properties of charmed baryons in nuclear
matter~\cite{Hosaka:2016ypm}. Meanwhile, there is a long history to
study the charmed
hypernuclei~\cite{Bhamathi:1981yu,Bando:1983yt,Starkov:1986ye}. The
conventional nuclei would become the charmed one when one or more
inner nucleons are replaced by the charmed baryons. The charmed
hypernuclei is an analogy of strange heypernuclei. The $Y_cN$
interaction is a cornerstone to understand the charmed hypernuclei.
Recently, the experimental proposals at
J-PARC~\cite{Noumi:2017sdz,Fujioka:2017gzp} and
FAIR~\cite{Wiedner:2011mf} on the charmed hypernuclei also inspired
numerous theoretical researches on the $Y_cN$ interactions and
charmed
hypernuclei~\cite{Cai:2003ce,Tsushima:2003dd,Kopeliovich:2007kd,Garcilazo:2015qha,Vidana:2019amb,Liu:2011xc,Maeda:2015hxa,Maeda:2018xcl,Huang:2013zva,Garcilazo:2019ryw,Miyamoto:2017tjs,Miyamoto:2017ynx,Haidenbauer:2017dua}.
See Refs.~\cite{Hosaka:2016ypm,Krein:2017usp} for recent reviews.

In Ref.~\cite{Liu:2011xc}, the OBE model was adopted to calculate
the $Y_cN$ potential. The molecular bound states of $\Lambda_cN$
were obtained with the essential couple-channel effect from
$\Sigma_cN$ and $\Sigma_c^*N$. In Ref.~\cite{Huang:2013zva}, the
$Y_cN$ and $Y_bN$ potentials were calculated within the framework of
the quark delocalization color screening model (QDCSM). The results
showed that the attraction between $N$ and $\Lambda_{c/b}$ is too
weak to form the bound states even with the couple-channel effect.
The $\Sigma_{c/b}N(^3S_1)$ resonance was obtained by coupling to the
$D$-wave $\Lambda_cN$ channel. In Ref.~\cite{Garcilazo:2019ryw}, a
constituent quark model was performed to study the $Y_cN$
interaction, which points to the soft interaction without two-body
bound states and weak couple-channel effect. In
Refs.~\cite{Maeda:2015hxa,Maeda:2018xcl}, a potential model ($Y_cN$
CTNN) combining the OBE model and quark model was constructed. The
model contains a long-range meson exchange part ($\pi$ and $\sigma$)
and a short distance quark exchange part. It is interesting that the
quark model~\cite{Huang:2013zva,Garcilazo:2019ryw} and the OBE
model~\cite{Liu:2011xc} present different pictures about $Y_cN$
interaction.

In Refs.~\cite{Miyamoto:2017tjs,Miyamoto:2017ynx}, the HAL QCD
collaboration presented the $\Lambda_cN$ potential and preliminary
$\Sigma_{c}N$ potential from the lattice QCD simulations. These
simulations from the first principle were performed in the quark
mass corresponding to the pion masses $m_{\pi}=410-700$ MeV. The
$\Lambda_cN$ interaction from lattice QCD was extrapolated to the
physical pion mass with the chiral effective field
theory~\cite{Haidenbauer:2017dua}. Rather than following the
standard Weinberg's power counting law, the authors omitted the
two-pion exchange contribution at the next-to-leading order and
included some high order contact terms. More refined calculation on
$Y_cN$ interaction with less model-dependent framework is needed.

In this work, we adopt the heavy baryon chiral perturbation theory
(HBChPT) to calculate the $\Sigma_{c}N$ interaction to the
next-to-leading order. We consider the leading order contact
interaction, one-pion-exchange contribution and the next-to-leading
order contact interaction and two-pion-exchange diagrams. We include
the $\Sigma_{c}^{(*)}$ as intermediate states in the loop diagrams.
With the analytical results, we extrapolate the phase shift from HAL
QCD to the physical pion mass and obtain the $\Sigma_{c}N$
interaction with quantum number $^3S_1(I=1/2)$. We also use a quark
model to relate the leading-order contact interaction of
$\Sigma_{c}N$ to those of $NN$ systems. In this approach, combining
the quark model and HBChPT, we give the numerical results of the
$S$-wave $\Sigma_{c}N$ interaction.

This paper is arranged as follows. In Sec.~\ref{sec:lag}, we
construct the Lagrangians. In Sec.~\ref{sec:ptl}, we calculate the
analytical results of the $\Sigma_{c}N$ interaction to the
next-to-leading order. We discuss the pion mass dependence of our
analytical results in Sec.~\ref{sec:pionmass}, and extrapolate the
lattice QCD results to the physical pion mass in Sec.~\ref{sec:num}.
We give a brief conclusion in Sec.~\ref{sec:conclud}. In
Appendix~\ref{app:qm}, we combine the quark model and analytical
results from HBChPT to give more predictions on the $\Sigma_{c}N$
interaction. In Appendix~\ref{app:super}, we present the integrals
used in our calculation.

\section{Effective Lagrangians and Weinberg's Formalism}\label{sec:lag}

We perform the chiral expansion in the framework of HBChPT. The
expansion is organized in powers of $\epsilon=q/\Lambda_{\chi }$,
where $q$ is either the momenta of Goldstone bosons or the residual
momenta of the matter fields. $\Lambda_{\chi}$ is the chiral
symmetry breaking scale. We calculate the scattering amplitude order
by order according to the power counting given by
Weinberg~\cite{Weinberg:1990rz,Weinberg:1991um}. In our calculation,
we include the $\Sigma_c^*$, the heavy quark spin symmetry partner
of $\Sigma_{c}$ as the intermediate states. In the analytical
calculations, we keep the mass splitting
$\delta_1=M_{\Sigma_{c}^*}-M_{\Sigma_c}$.

The amplitudes of box diagrams would blow up in the heavy baryon
limit, which would also been amplified to destroy the power counting
even if we include the kinetic terms of the heavy baryon. Thus, we
adopt the Weinberg's formalism to deal with this
problem~\cite{Weinberg:1990rz,Weinberg:1991um,Kaiser:1997mw,Meng:2019ilv,Wang:2019ato}.
We subtract the two particle reducible (2PR) contribution in the box
diagrams, which could be generated by iterating the
one-pion-exchange diagrams. The remaining two particle irreducible
(2PIR) part obeys the power counting, which is treated as the kernel
of the Lippmann-Schwinger equation or Schr\"{o}dinger equation.

 We introduce the pion field and the involved building blocks as follows,
\begin{eqnarray}
&&\phi=\sqrt{2}\left(\begin{array}{cc}
\frac{\pi^{0}}{\sqrt{2}} & \pi^{+}\\
\pi^{-} & -\frac{\pi^{0}}{\sqrt{2}}
\end{array}\right), U=u^{2}=\exp\left(i\frac{\phi(x)}{F}\right),\\
&&\Gamma_{\mu}=\frac{1}{2}[u^{\dagger},\partial_{\mu}u],\quad
u_{\mu}=\frac{i}{2}\{u^{\dagger},\partial_{\mu}u\},
\end{eqnarray}
where the $\Gamma_{\mu}$ and $u_{\mu}$ are the chiral connection and
axial-vector current, respectively. $F$ is the pion decay constant.

The multiplets of spin-$1\over 2$ $\Sigma_c$ and spin-$3\over 2$
$\Sigma_c^*$ are represented as
\begin{eqnarray}
\Sigma_{c}=\left(\begin{array}{cc}
\Sigma_{c}^{++} & \frac{\Sigma_{c}^{+}}{\sqrt{2}}\\
\frac{\Sigma_{c}^{+}}{\sqrt{2}} & \Sigma_{c}^{0}
\end{array}\right),\quad\Sigma_{c}^{*\mu}=\left(\begin{array}{cc}
\Sigma_{c}^{*++} & \frac{\Sigma_{c}^{*+}}{\sqrt{2}}\\
\frac{\Sigma_{c}^{*+}}{\sqrt{2}} & \Sigma_{c}^{*0}
\end{array}\right)^{\mu}.
\end{eqnarray}
The leading order Lagrangians associated with $\Sigma_c^{(*)}$ are
constructed as
\begin{eqnarray}
{\cal L}_{\Sigma_{c}^{(*)}\phi}^{(0)}&=&\text{Tr}[\bar{\Sigma}_{c}(i\slashed{D}-M_{\Sigma_{c}})\Sigma_{c}]+ \text{Tr}[\bar{\Sigma}_{c}^{*\mu}[-g_{\mu\nu}(i\slashed{D}-M_{\Sigma_{c}^{*}})\nonumber \\
&&+i(\gamma_{\mu}D_{\nu}+\gamma_{\nu}D_{\mu})-\gamma_{\mu}(i\slashed{D}+M_{\Sigma_{c}^{*}})\gamma_{\nu}]\Sigma_{c}^{*\nu}]\nonumber \\
&&+g_1\text{Tr}[\bar{\Sigma}_{c}\gamma^{\mu}\gamma_{5}u_{\mu}\Sigma_{c}]+g_3\text{Tr}[\bar{\Sigma}_{c}^{*\mu}u_{\mu}\Sigma_{c}+\text{H.c.}]\nonumber \\
&&+g_5\text{Tr}[\bar{\Sigma}_{c}^{*\nu}\gamma^{\mu}\gamma_{5}u_{\mu}\Sigma_{c\nu}^{*}],
\label{lag:siglo}
\end{eqnarray}
where $\text{Tr}[...]$ represents the trace in the flavor space.
$M_{\Sigma_{c}}$ and $M_{\Sigma_{c}^*}$ denote the masses of
$\Sigma_c$ and $\Sigma_c^\ast$, respectively. The covariant
derivative is defined as
$D_{\mu}\Sigma_{c}^{(*)}=\partial_{\mu}\Sigma_{c}^{(*)}+\Gamma_{\mu}\Sigma_{c}^{(*)}+\Sigma_{c}\Gamma_{\mu}^{(*)T}$.
The $g_1$, $g_3$ and $g_5$ are axial coupling constants. We can
define the superfield to set up the relation of $\Sigma_{c}$ and
$\Sigma_c^*$ in the heavy quark limit,
\begin{eqnarray}
\psi^{\mu}  &=&\mathcal{B}^{*\mu}-\sqrt{\frac{1}{3}}(\gamma^{\mu}+v^{\mu})\gamma^{5}\mathcal{B},\nonumber \\
\bar{\text{\ensuremath{\psi}}}^{\mu}
&=&\bar{\mathcal{B}}^{*\mu}+\sqrt{\frac{1}{3}}\bar{\mathcal{B}}\gamma^{5}(\gamma^{\mu}+v^{\mu}),
\end{eqnarray}
where $\mathcal{B}^{(*)}$ are the $\Sigma_c^{(*)}$ fields after
heavy baryon reduction. The Lagrangians can be rewritten as a more
compact form with the superfield,
\begin{eqnarray}\label{eq:SigmaLO}
{\cal L}_{\Sigma_{c}\phi}^{(0)}=&-&\text{Tr}[\bar{\psi}^{\mu}iv\cdot D\psi_{\mu}]+ig_{a}\epsilon_{\mu\nu\rho\sigma}\text{Tr}[\bar{\psi}^{\mu}u^{\rho}v^{\sigma}\psi^{\nu}]\nonumber \\
&+&
i\frac{\delta_{1}}{2}\text{Tr}[\bar{\psi}^{\mu}\sigma_{\mu\nu}\psi^{\nu}].
\label{lag:siglohq}
\end{eqnarray}
where the third term represents the heavy quark spin symmetry
violation effect. $\delta_{1} =M_{\Sigma_{c}^{*}}-M_{\Sigma_{c}}$ is
the mass splitting between $\Sigma_{c}$ and $\Sigma_{c}^*$.
Comparing Eq.~(\ref{lag:siglohq}) with Eq.~(\ref{lag:siglo}), one
can easily get
\begin{equation}
g_{1}=-\frac{2}{3}g_{a},\quad g_{3}=-\sqrt{\frac{1}{3}}g_{a},\quad
g_{5}=g_{a}.
\end{equation}

The leading order Lagrangians for nucleons can be constructed as
\begin{eqnarray}
{\cal L}_{N\phi}^{(0)}
=\bar{N}(i\slashed{\mathcal{D}}-M_{N})N+g_{A}\bar{N}\gamma^{\mu}\gamma_{5}u_{\mu}N,
\end{eqnarray}
where $N=(p,n)^T$ is the nucleon isospin doublet. $M_N$ and $g_A$
are the nucleon mass and axial coupling constant, respectively. The
covariant derivative is define as
$\mathcal{D}_{\mu}=\partial_{\mu}+\Gamma_{\mu}$.

At the leading order, the contact terms also contribute to the
$\Sigma_cN$ interaction. The independent Lagrangians read
\begin{eqnarray}
{\cal L}^{(0)}_{\text{contact}} &=&C_{3}\bar{N}N\text{Tr}(\bar{\psi}^{\mu}\psi_{\mu})+C_{4}(\bar{N}\bm{\tau}N)\cdot\text{Tr}(\bar{\psi}^{\mu}\bm{\tau}\psi_{\mu}) \nonumber\\
&+&i\frac{3}{2}\tilde{C}_{3}(\bar{N}\sigma_{\mu\nu}N)\text{Tr}(\bar{\psi}^{\mu}\psi^{\nu})\nonumber\\
&+&i\frac{3}{2}\tilde{C}_{4}(\bar{N}\bm{\tau}\sigma_{\mu\nu}N)\cdot\text{Tr}(\bar{\psi}^{\mu}\bm{\tau}\psi^{\nu}),
\label{lag:loctc}
\end{eqnarray}
where $\bm{\tau}$ is the Pauli matrix in the isospin space. $C_3$,
$\tilde{C}_3$, $C_4$ and $\tilde{C}_4$ are the low energy constants
(LECs).

\section{effective potentials}\label{sec:ptl}

We calculate the  $\Sigma_{c}N$ effective potential
$\mathcal{V}(\bm{q})$ in the momentum space to the next-to-leading
order. The chiral dynamics is essentially the interplay of the light
degrees of freedom. For the $\Sigma_{c}^{(*)}N$ system, the elements
of the interaction are the spin triplet light diquark in
$\Sigma_{c}^{(*)}$ and $N$. In the calculation, we include both
$\Sigma_c$ and $\Sigma_c^*$ as the intermediate states to ensure the
whole spin triplet light diquark is considered. The $\Lambda_c$ and
$\Sigma_{c}^{(*)}$ belong to the different isospin multiplets. The
$\Lambda_c$ plays the similar role as the $\Delta$ in the $NN$
system~\cite{Krebs:2007rh}. In this work, we include neither the
$\Delta$ nor the $\Lambda_c$ as intermediate states.

 \begin{figure}[!htp]
    \centering
    \includegraphics[width=0.3\textwidth]{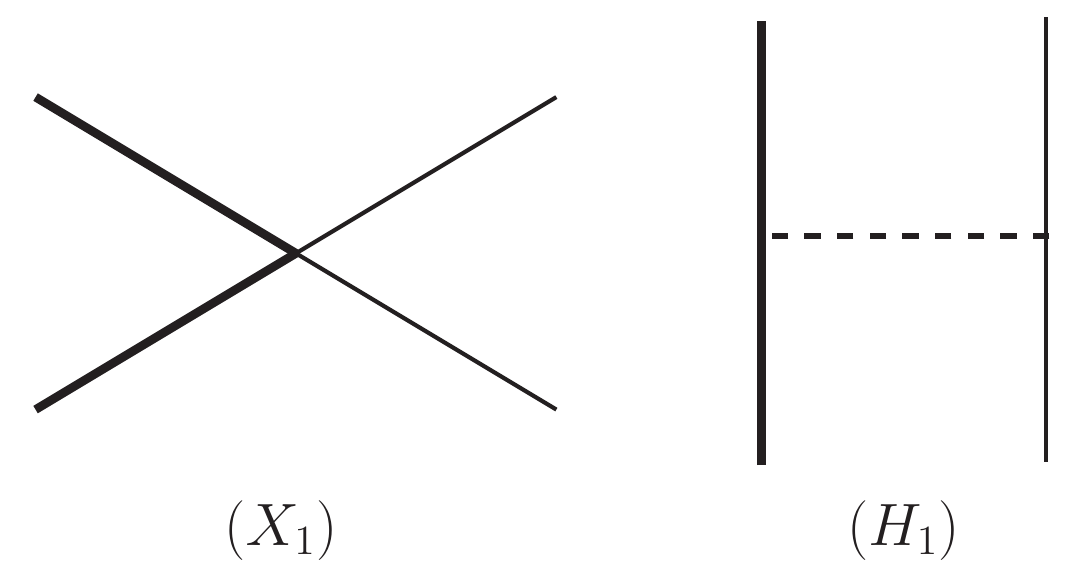}
    \caption{The leading order diagrams for the $\Sigma_cN$
        system. At this order, the contact term $(X_{1})$ and the
        one-pion-exchange diagram $(H_{1})$ contribute to the effective
        potential. The solid, thick solid and dashed lines represent the $N$, $\Sigma_c$ and pion, respectively.}\label{fig:leading}
\end{figure}

The leading order $\mathcal{O}(\epsilon^0)$ potential of $\Sigma_cN$
arises from the tree level contact and one-pion-exchange diagrams in
Fig.~\ref{fig:leading}. The leading order potential reads
\begin{eqnarray}
&&\mathcal{V}_{X_1}^{(0)}=C_{3}+2C_{4}\bm{I}_{1}\cdot\bm{I}_{2}+\left(\tilde{C}_{3}+2\tilde{C}_{4}\bm{I}_{1}\cdot\bm{I}_{2}\right)\bm{\sigma}_{1}\cdot\bm{\sigma}_{2}, \nonumber \\
&&{\cal
V}_{H_1}^{(0)}=-\frac{g_{A}g_{1}}{2F^{2}}(\bm{I}_{1}\cdot\bm{I}_{2})\frac{(\bm{\sigma}_{1}\cdot\bm{q})(\bm{\sigma}_{2}\cdot\bm{q})}{\bm{q}^{2}+m^{2}},
\label{eq:vlo}
\end{eqnarray}
where the subscript $X_1$ and $H_1$ denote the label of Feynman
diagrams. The $m$ is the pion mass. $\bm{I}_1\cdot \bm{I}_2$ and
$\bm{\sigma}_{1}\cdot \bm{\sigma}_{2}$ are the isospin-isospin and
spin-spin operators, respectively. Their matrix elements are
calculated as
\begin{eqnarray}
\langle\bm{I}_1\cdot \bm{I}_2\rangle&=&{1\over 2}[I(I+1)-I_1(I_1+1)-I_2(I_2+1)],\nonumber\\
\langle\bm{\sigma}_1\cdot
\bm{\sigma}_2\rangle&=&2[S(S+1)-S_1(S_1+1)-S_2(S_2+1)].\nonumber
\end{eqnarray}
Since we only focus on the $S-$wave $\Sigma_{c}N$ system, we can
take the folllowing replacement,
\begin{equation}
q^iq^j \rightarrowtail {1\over {3}}\bm{q^2}\delta^{ij}.
\end{equation}

The next-to-leading order potential arises from the
$\mathcal{O}(\epsilon^2)$ tree level contact term, the
two-pion-exchange interaction and the renormalization of the
$\mathcal{O}(\epsilon^0)$ diagrams. One can construct the general
contact interaction~\cite{Epelbaum:2004fk}. In this work, we only
concentrate on the $S$-wave interaction and omit the recoil terms.
Thus, the next-to-leading order contact potential reads
\begin{eqnarray}
\mathcal{V}_{X_1}^{(2)}&=&\left(C_{5}+C_{6}\bm{I}_{1}\cdot\bm{I}_{2}\right)\bm{q}^{2} \nonumber\\
&&+\left(\tilde{C}_{5}+\tilde{C}_{6}\bm{I}_{1}\cdot\bm{I}_{2}\right)\bm{q}^{2}(\bm{\sigma}_{1}\cdot\bm{\sigma}_{2}),
\label{eq:vnlo}
\end{eqnarray}
where $C_5$, $C_6$, $\tilde{C}_5$ and $\tilde{C}_6$ are LECs in the
$\mathcal{O}(\epsilon^2)$ contact Lagrangians.
\begin{figure*}[!htp]
    \centering
    \includegraphics[width=0.85\textwidth]{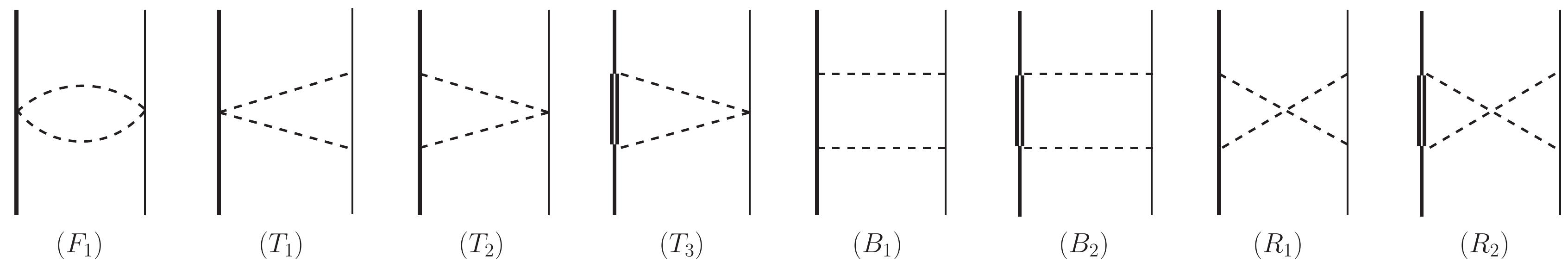}
    \caption{The two-pion-exchange diagrams for the $\Sigma_cN$ system.
        There is one football diagram ($F_{1}$), three triangle diagrams
        ($T_{1,2,3}$), two box diagrams ($B_{1,2}$) and two crossed box
        diagrams ($R_{1,2}$). The solid, thick solid, double
        thick solid and dashed lines represent the $N$, $\Sigma_c$, $\Sigma_c^*$ and pion fields, respectively.}\label{fig:2phi}
\end{figure*}

At the next-to-leading order, the two-pion-exchange contributions
read
\begin{widetext}
\begin{eqnarray}
{\cal V}^{(2)}_{F_{1}}&=&\frac{1}{F^{4}}(\bm{I}_{1}\cdot\bm{I}_{2})J_{22}^{F},\\
{\cal V}^{(2)}_{T_{1}}&=&\frac{-4g_{A}^{2}}{F_{0}^{4}}(\bm{I}_{1}\cdot\bm{I}_{2})\left[\frac{1}{4}\bm{q^{2}}(J_{24}^{T}+J_{33}^{T})+\frac{1-d}{4}J_{34}^{T}\right](0),\\
{\cal V}^{(2)}_{T_{2}}&=&\frac{-g_{1}^{2}}{F_{0}^{4}}(\bm{I}_{1}\cdot\bm{I}_{2})\left[\frac{1}{4}\bm{q^{2}}(J_{24}^{T}+J_{33}^{T})+\frac{1-d}{4}J_{34}^{T}\right](0),\\
{\cal V}^{(2)}_{T_{3}}&=&\frac{g_{3}^{2}}{4F^{4}}(\bm{I}_{1}\cdot\bm{I}_{2})\left[J_{34}^{T}(d-2)+(J_{33}^{T}+J_{24}^{T})\bm{q^{2}}\frac{2-d}{d-1}\right](-\delta_{1}),\\
{\cal V}^{(2)}_{B_{1}}&=&\frac{g_{A}^{2}g_{1}^{2}}{F^{4}}2(1-\bm{I_{1}\cdot I_{2}})\Bigg\{ J_{41}^{B}\left[\left(\frac{1-d}{4}\right)^{2}+\frac{3}{8}\right]-(J_{42}^{B}+J_{31}^{B})\frac{1+d}{8}\bm{q^{2}}-J_{21}^{B}\frac{1}{16}\bm{q^{2}}\nonumber\\
&+&(J_{43}^{B}+2J_{32}^{B}+J_{22}^{B})\frac{1}{16}\bm{q^{4}}+[(\bm{\sigma}_{1}\cdot\bm{\sigma}_{2})\bm{q}^{2}-\mathbb{T}]J_{21}^{B}\bm{q}^{2}\Bigg\} (0,0),\\
\mathcal{V}^{(2)}_{B_{2}}&=&-\frac{g_{A}^{2}g_{3}^{2}}{8F^{4}}(1-\bm{I_{1}\cdot I_{2}})\Bigg\{J_{41}^{B}\left[(1-d)d+\frac{6}{(d-1)}\right]+(J_{42}^{B}+J_{31}^{B})\frac{2(d-2)(d+1)}{(d-1)}\bm{q}^{2}+J_{21}^{B}\frac{d-2}{d-1}\bm{q}^{2}\\
&+&(J_{43}^{B}+2J_{32}^{B}+J_{22}^{B})\frac{2-d}{d-1}\bm{q}^{4}+[(\bm{\sigma}_{1}\cdot\bm{\sigma}_{2})\bm{q}^{2}-\mathbb{T}]J_{21}^{B}\frac{\bm{q}^2}{d-1}\Bigg\}(-\delta_{1},0),\nonumber \\
\mathcal{V}^{(2)}_{R_i}&=&\mathcal{V}^{(2)}_{B_i}|_{J_x^B\to
J_x^R,~\bm{I}_1\cdot\bm{I}_2\to
-\bm{I}_1\cdot\bm{I}_2,~\bm{\sigma}_1\cdot\bm{\sigma}_2\to
-\bm{\sigma}_1\cdot\bm{\sigma}_2},
\end{eqnarray}
\end{widetext}
where $d$ is the dimension in the dimensional regularization.
$\mathbb{T}=(\bm{\sigma}_{1}\cdot
\bm{q})(\bm{\sigma}_2\cdot\bm{q})$. $J^{X}_{ij}$ are the loop
integrals defined in the Appendix~\ref{app:super}. They are the
functions of $m$, $\bm{q}^2$ and mass splitting. For conciseness, we
omit the $m$ and $\bm{q}^2$ and keep the specific mass splitting at
the end of every expression.

The vertex renormalization, mass renormalization and wave function
renormalization of the leading order diagrams will also contribute
to the next-to-leading order potential. The renormalizations of the
one-pion-exchange diagram and the leading order contact term are
presented in Figs.~\ref{fig:onepioncrrct} and
\ref{fig:contactcrrct}, respectively. When we calculate the
$\Sigma_cN$ potential at the physical pion mass, these
renormalization effects can be included by adopting the physical
hadron masses and coupling constants in the tree level results. When
we vary the pion mass, these renormalization effects will induce the
extra pion mass dependence. We will discuss these renormalization
diagrams in the next section.

\section{pion mass dependence}\label{sec:pionmass}

The $\Sigma_cN$ potentials depend on the pion mass either explicitly
in expressions of Sec.~\ref{sec:ptl}, or implicitly through the
decay constants, the LECs, the mass of baryons and the
renormalization of wave functions.

\begin{figure}[!htp]
    \centering
    \includegraphics[width=0.45\textwidth]{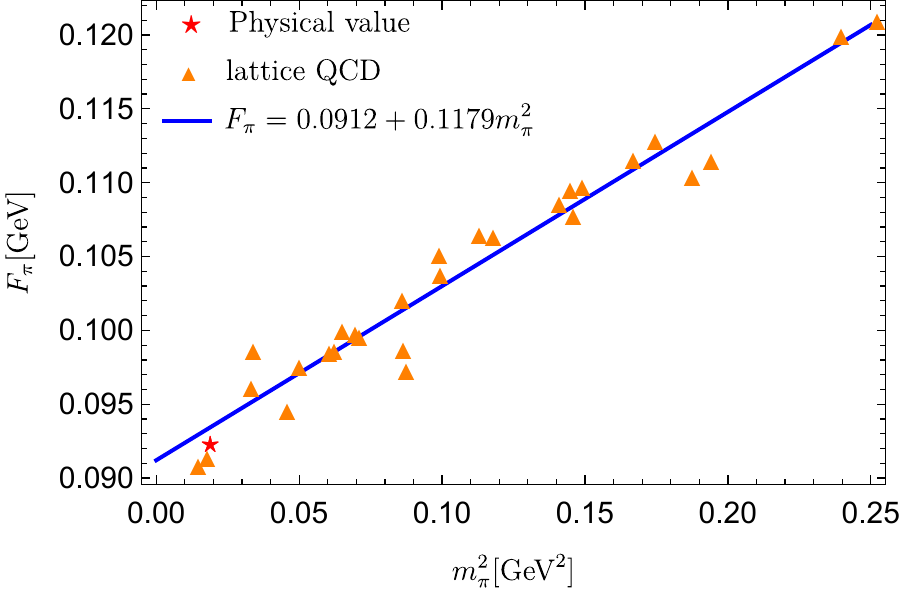}
    \caption{The $m_{\pi}$-dependence of pion decay constant from Ref.~\cite{Durr:2013goa} and a linear fit in this work. For the lattice QCD results, we only give the central value here.}\label{fig:fpi}
\end{figure}

The $m_{\pi}$-dependence for the pion decay constant can be obtained
in the framework of ChPT. It was shown that the next-to-leading
SU(2) results can not fit the $m_{\pi}$-dependent decay constant
from lattice QCD well in FIG.~3 of Ref.~\cite{Durr:2013goa}. As
shown in FIG.~5 of Ref.~\cite{Durr:2013goa}, one needs to calculate
at least to the next-to-next-to-leading order to depict the
$m_{\pi}$-dependence, which is beyond the precision of this work.
Rather than going to higher order calculation, we fit the lattice
results with a linear function of $m_{\pi}^2$. The result is shown
in Fig.~\ref{fig:fpi}. The simple linear function can depict the
$m_{\pi}$-dependence rather well at least when the pion mass is less
than $500$ MeV.

The renormalization of one-pion-exchange diagram would contribute to
the next-to-leading order amplitude,
\begin{equation}
\mathcal{A}_{\mathrm{ope}}^{r}=(1+\delta Z_{\pi}+\delta Z_{N}+\delta
Z_{\Sigma_c})\mathcal{A}_{\mathrm{ope}}(g_i^r,m_r),
\end{equation}
where $g_i^r$ and the $m_r$ denote the renormalized axial coupling
constants and hadron masses. The $\delta Z_{\pi}$, $\delta Z_{N}$
and $\delta Z_{\Sigma_c}$ arise from the wave function
renormalization  of $\pi$, $N$ and $\Sigma_{c}$, respectively.

\begin{figure*}[!htp]
    \centering
    \includegraphics[width=0.8\textwidth]{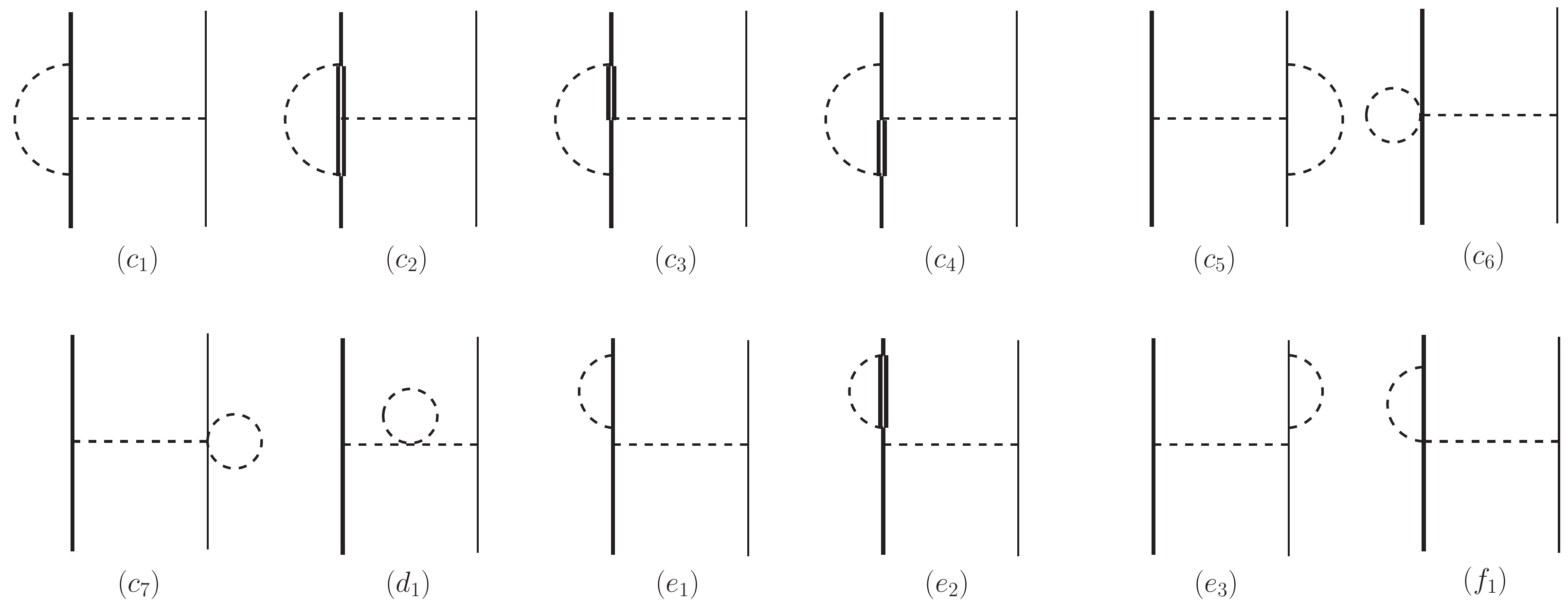}
    \caption{The renormalization diagrams for the one-pion-exchange interaction. The $(c_1)$-$(c_7)$ diagrams renormalize the axial coupling vertices. The $(d_1)$ diagram renormalize the pion field. The $(e_1)$ and $(e_2)$ renormalize the $\Sigma_c$ field. The $(e_3)$ renormalize the $N$ field. The $(f_1)$ diagram vanishes. Notations are the same as those in Fig.~\ref{fig:2phi}.}\label{fig:onepioncrrct}
\end{figure*}

Up to the next-to-leading order, the $m_{\pi}$-dependence of
renormalized $g^r_1$ and $g^r_A$ can also be obtained by calculating
the vertex renormalization diagrams $(c_1)$$-$$(c_7)$ in
Fig.~\ref{fig:onepioncrrct}. The results read
\begin{eqnarray}
&&g_{A}^{r}=g_{A}(1+\delta Z_{g_{A}}),\quad g_{1}^{r}=g_{1}(1+\delta Z_{g_{1}}),\\
&&\delta Z_{g_{A}}=-\frac{J_{0}^{c}}{3F^{2}}-\frac{d-3}{4}\frac{g_{A}^{2}}{F^{2}}J_{22}^{g}(0,0), \\
&& \delta Z_{g_{1}}=-\frac{J_{0}^{c}}{3F^{2}}+\frac{d-3}{4}\frac{g_{1}^{2}}{F^{2}}J_{22}^{g}(0,0), \nonumber\\
&&+\left[(d-3)+\left(\frac{d-3}{d-1}\right)^{2}\right]\frac{g_{3}^{2}g_{5}}{4F^{2}g_{1}}J_{22}^{g}(-\delta_{1},-\delta_{1}) \nonumber\\
&&+\frac{g_{3}^{2}}{F^{2}}\frac{2-d}{d-1}J_{22}^{g}(-\delta_{1},0).
\end{eqnarray}
The lattice QCD simulation in
Refs.~\cite{Procura:2006gq,Detmold:2012ge,Alexandrou:2016xok} showed
that the $g_A^r$ and $g_1^r$ are insensitive to the pion mass. Thus,
we keep the axial coupling constants $g_1$ and $g_A$ invariant with
the change of the pion mass.

The field renormalization diagrams $(d_1)$, $(e_1)$, $(e_2)$ and
$(e_3)$ in Fig.~\ref{fig:onepioncrrct} would give the wave function
renormalization and mass renormalization. The mass renormalization
effects are included by adopting the $m_{\pi}$-dependent hadron
masses in Table~\ref{tab:mass}. The wave function renormalizations
read
\begin{eqnarray}
\delta Z_{\pi}&=&\frac{2J_{0}^{c}}{3F^{2}}, \label{eq:zpi}\\
\delta Z_{N}&=&\frac{g_{A}^{2}}{F^{2}}\frac{3(1-d)}{4}\frac{\partial J_{22}^{a}(\omega)}{\partial\omega}|_{\omega=0}, \label{eq:zn} \\
\delta Z_{\Sigma_{c}}&=&\frac{g_{1}^{2}}{F^{2}}\frac{1-d}{2}\frac{\partial J_{22}^{a}(\omega)}{\partial\omega}|_{\omega=0}\nonumber\\
&+&\frac{g_{3}^{2}}{F^{2}}\frac{2-d}{2}J_{22}^{a}\frac{\partial
J_{22}^{a}(\omega)}{\partial\omega}|_{\omega=-\delta_{1}}.
\label{eq:zsig}
\end{eqnarray}
Eqs.~\eqref{eq:zpi}-\eqref{eq:zsig} would bring extra
$m_{\pi}$-dependence to the potential.

\begin{table}
    \caption{Hadron masses at different pion mass (in units of MeV)~\cite{Miyamoto:2017ynx}. }\label{tab:mass}
    \setlength{\tabcolsep}{5.8mm}
    \begin{tabular}{c|cccc}
        \toprule[0.5pt]\toprule[0.5pt]
        $m_{\pi}$ & $m_{N}$ & $m_{\Sigma_{c}}$ & $m_{\Sigma_{c}^{*}}$ & $\delta_1$\tabularnewline
        \midrule[0.5pt]
        139 & 938 & 2454 & 2518 & 64\tabularnewline
        412 & 1215 & 2575 & 2661 & 86\tabularnewline
        570 & 1399 & 2674 & 2763 & 89\tabularnewline
        702 & 1581 & 2780 & 2866 & 86\tabularnewline
        \bottomrule[0.5pt]\bottomrule[0.5pt]
    \end{tabular}
\end{table}

\begin{figure}[!htp]
    \centering
    \includegraphics[width=0.49\textwidth]{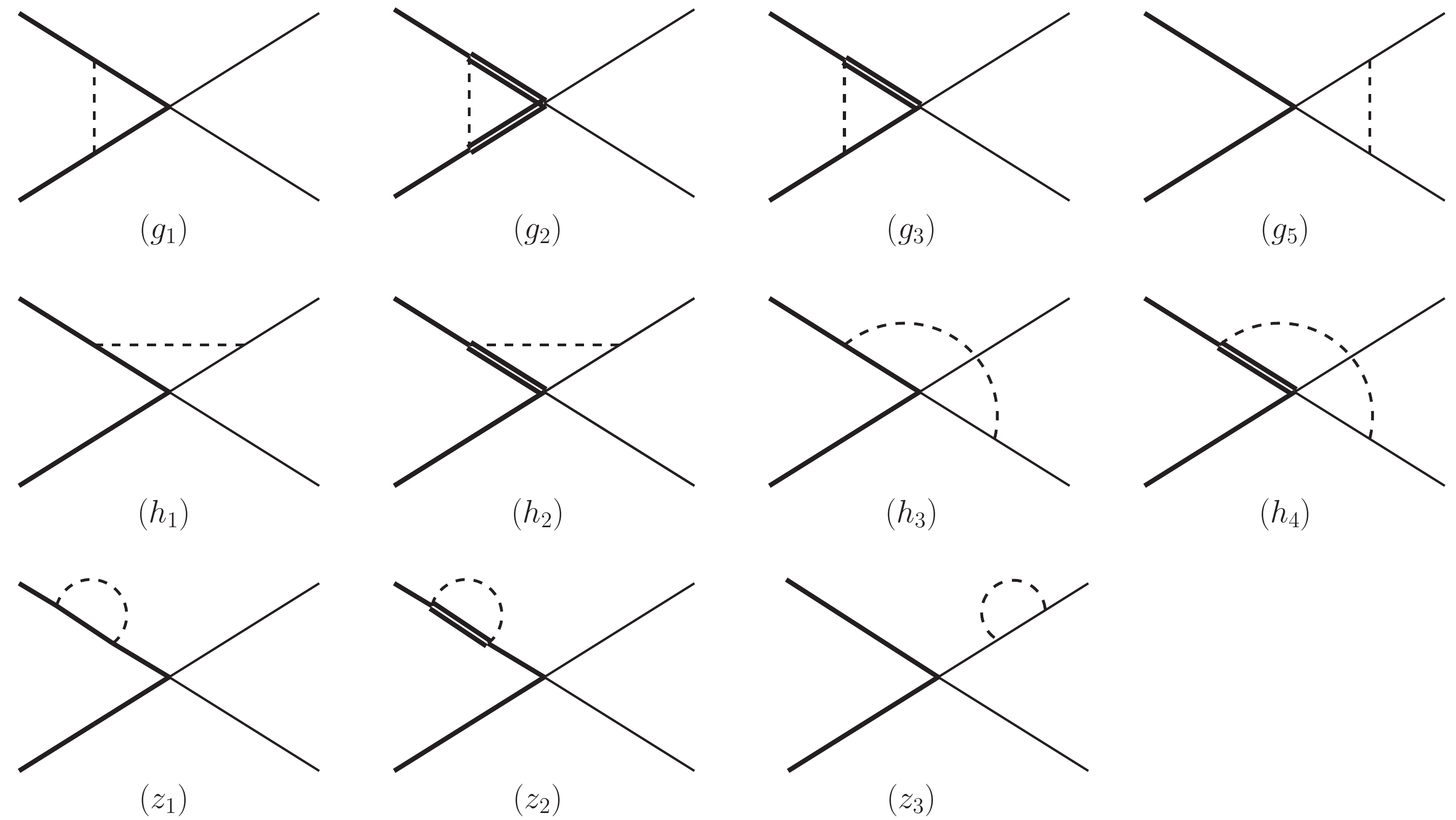}
    \caption{The renormalization diagrams for contact interaction. The $(g_i)$ and $(h_i)$ diagrams are the contact vertices renormalization. The $(z_i)$ diagrams are the wave function renormalization. Notations are the same as those in Fig.~\ref{fig:2phi}.}\label{fig:contactcrrct}
\end{figure}

In Fig.~\ref{fig:contactcrrct}, we present the Feynman diagrams
contributing to the renormalization of contact terms. The $(g_i)$
and $(h_i)$ diagrams would renormalize the leading order contact
terms,
\begin{eqnarray}
C_i^r=C_i+\delta C_i.
\end{eqnarray}
In the heavy quark limit and with $d\to4$, $\delta C_i$ read
\begin{eqnarray}
\delta C_{3}&=&\frac{1}{F^{2}}J_{22}^{g}\left[-C_{3}g_{a}^{2}-C_{3}\frac{9}{4}g_{A}^{2}\right],\nonumber\\
\delta\tilde{C}_{3}&=&\frac{1}{F^{2}}J_{22}^{g}\left[-\frac{13}{36}\tilde{C}_{3}g_{a}^{2}+\frac{3}{4}\tilde{C}_{3}g_{A}^{2}+(\tilde{C_{3}}+2\tilde{C}_{4})g_{a}g_{A}\right], \nonumber\\
\delta C_{4}    &=&\frac{1}{F^{2}}J_{22}^{g}\left[-\frac{1}{2}C_{4}g_{a}^{2}-\frac{9}{8}C_{4}g_{A}^{2}+3\tilde{C}_{4}g_{a}g_{A}\right],\nonumber\\
\delta\tilde{C}_{4}
&=&\frac{1}{F^{2}}J_{22}^{g}\left[-\frac{13}{72}\tilde{C}_{4}g_{a}^{2}+\frac{3}{8}\tilde{C}_{4}g_{A}^{2}+\frac{2}{3}C_{4}g_{a}g_{A}\right],
\end{eqnarray}
where the $J^g_{22}$ has the finite terms,
\begin{equation}
J^g_{22}=-{1\over
16\pi^2(d-1)}\left[2m^{2}+3m^{2}\ln\frac{m^{2}}{\lambda^{2}}\right].
\end{equation}
where $\lambda$ is the chiral symmetry breaking scale
$\Lambda_{\chi}$. When the spin and the isospin of $\Sigma_cN$ are
fixed, the $m_{\pi}$-dependent contact interaction can be
reparameterized as
\begin{eqnarray}\label{eq:repara}
\mathcal{V}_{\mathrm{contact}}&=&c_{1}(1+\delta Z_{N}
+\delta Z_{\Sigma_{c}}) \nonumber\\
&+&\tilde{c}_{1}\frac{m^{2}}{F^{2}}\left(2+3\ln\frac{m^{2}}{\lambda^{2}}\right)+c_2\bm{q}^2+c_3
m^2,
\end{eqnarray}
where the $c_1$ terms denote the leading order interaction and its
wave function renormalization. The $\tilde{c}_1$ terms represent the
renormalization of the leading order contact coupling constants. The
$c_2$ and $c_3$ terms arise from the next-to-leading order tree
diagrams.

\section{Numerical results}\label{sec:num}

We keep the mass splitting between $\Sigma_{c}^*$ and $\Sigma_{c}$
when we calculate the analytical results. In the real physical
world, the mass splitting $\delta_1\simeq64$ MeV is of the same
order as the physical pion mass. Thus, we can adopt the
small-scale-expansion and treat the $\delta_1$ as another small
scale like the pion mass~\cite{Hemmert:1997ye}. However, when we go
to the world with much heavier pion as shown in
Table~\ref{tab:mass}, the mass splitting become much less than the
pion mass. The small-scale-expansion for $\delta_{1}$ does not work
any more. Thus, in this work, we take the heavy quark limit and omit
the mass splitting $\delta_1$. The 2PR contribution in the box
diagram ($B_2$) of Fig.~\ref{fig:2phi} and diagram ($h_2$) of
Fig.~\ref{fig:contactcrrct} should be subtracted, which is different
from our previous works~\cite{Meng:2019ilv,Wang:2019ato}.

In this work, we choose the $m_{\pi}$-dependent chiral symmetry
breaking scale $\Lambda_{\chi }=4\pi F(m_{\pi})$. An alternative
approach to the $m_{\pi}$-dependent $\Lambda_{\chi }$ is the mass of
$\rho$ meson~\cite{Leinweber:2001ac,Werner:2019hxc}, which brings
the slight divergence from the former one. To ensure the good chiral
convergence, we only adopt the lattice QCD results with
$m_{\pi}\approx 410$ MeV and $570$ MeV in
Ref.~\cite{Miyamoto:2017ynx}.

With the potential in the momentum space, the potential in the
coordinate space reads
\begin{equation}
V(r)=\frac{1}{(2\pi)^{3}}\int
d^{3}\bm{q}\,e^{i\bm{q}\cdot\bm{r}}{\cal V}(\bm{q}){\cal F}(\bm{q}).
\end{equation}
where ${\cal F}(\bm{q})=\exp(-\bm{q}^{2n}/\Lambda^{2n})$ is the
regulator to suppress the contribution of the high
momentum~\cite{Ordonez:1995rz,Machleidt:2011zz,Ren:2016jna}. In this
work, we choose two different regulators. In scenario I, we adopt
$n=1$ and $\Lambda=0.8$ GeV. In scenario II, we adopt $n=2$ and
$\Lambda=0.5$ GeV.

We then calculate the phase shift from the potential, which is the
physical observable. The partial wave Lippmann-Schwinger equation
reads~\cite{Stern:1979zj}
\begin{eqnarray}
&&\psi_{l}^{P}(k,r)=u_{l}(kr)+2\mu\int_{0}^{\infty}dr'G_{l}^{P}(k;r,r')V(r')\psi_{l}^{P}(k,r'),\nonumber\\
&&G_{l}^{P}=k^{-1}u_{l}(kr_{<})v_{l}(kr_{>}),
\end{eqnarray}
where $u_{l}(kr)$ and $v_l(kr)$ are Riccati-Bessel function and
Riccati-Neumann function, respectively. $r_{>}=\max\{r,r'\}$ and
$r_{<}=\min\{r,r'\}$. The superscript ``$P$'' denotes the Green
function and the wave functions are the principal-value ones. The
$K$-matrices and the phase shifts can be obtained by
\begin{eqnarray}
K_{l}=\tan\delta_{l}=-2\mu
k^{-1}\int_{0}^{\infty}dru_{l}(kr)V(r)\psi_{l}^{P}(k,r).
\end{eqnarray}
The scattering length could be obtained by performing the
effective-range expansion. For the $S$-wave, the expansion reads
\begin{equation}
k\cot\delta=-\frac{1}{a_{s}}+\frac{1}{2}k^{2}r_{e}+\cdots,
\end{equation}
where $a_s$ and $r_e$ are the scattering length and the effective
range, respectively.

\begin{figure*}[!htp]
    \centering
    \includegraphics[width=0.3\textwidth]{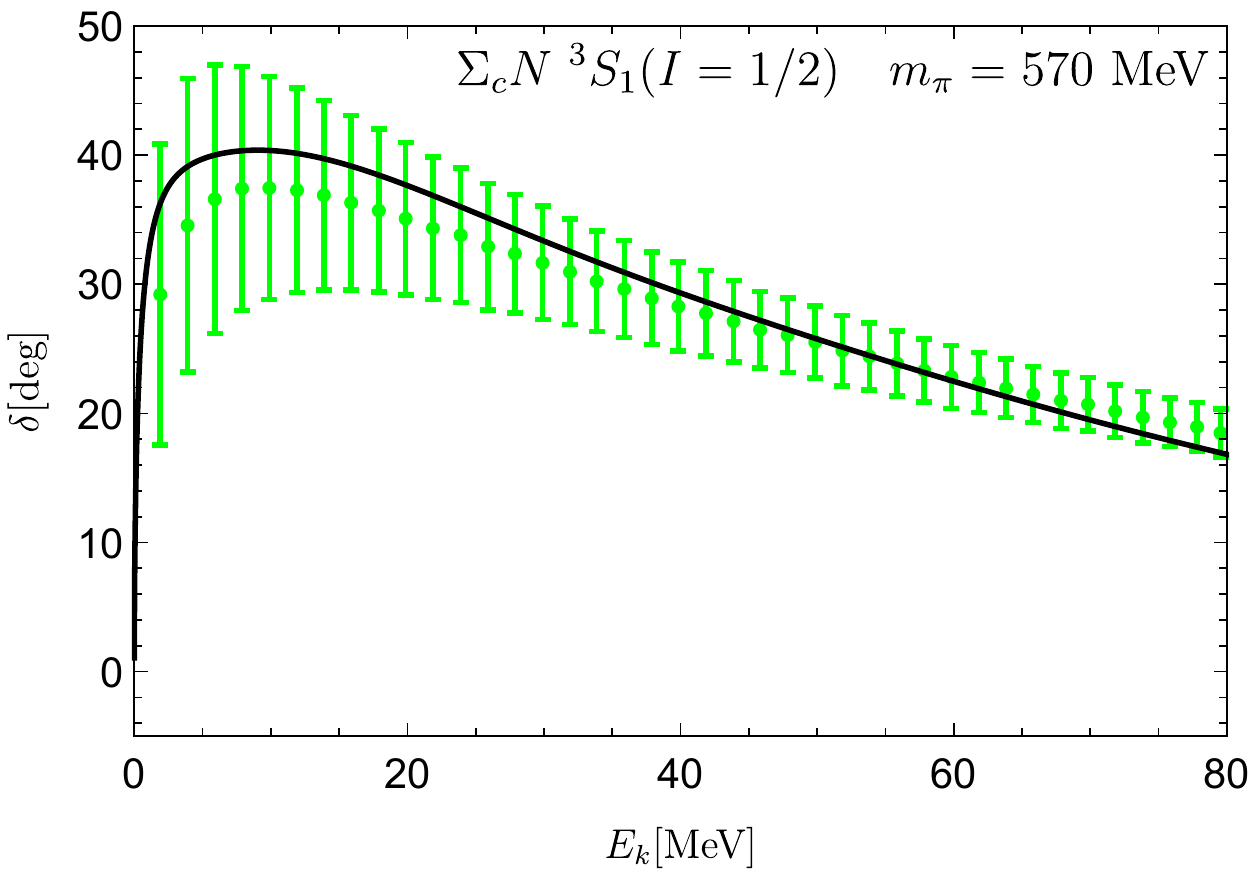}
    \includegraphics[width=0.3\textwidth]{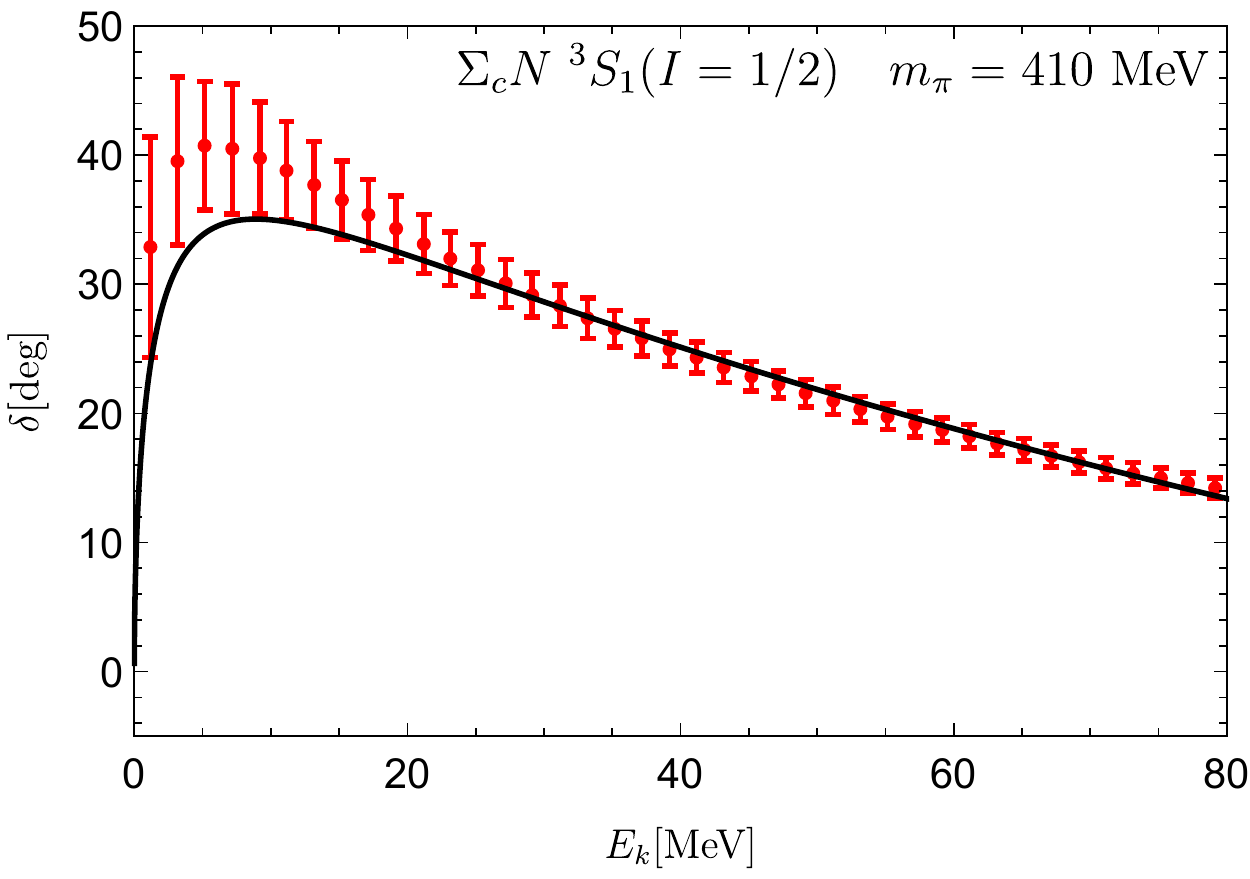}
    \includegraphics[width=0.3\textwidth]{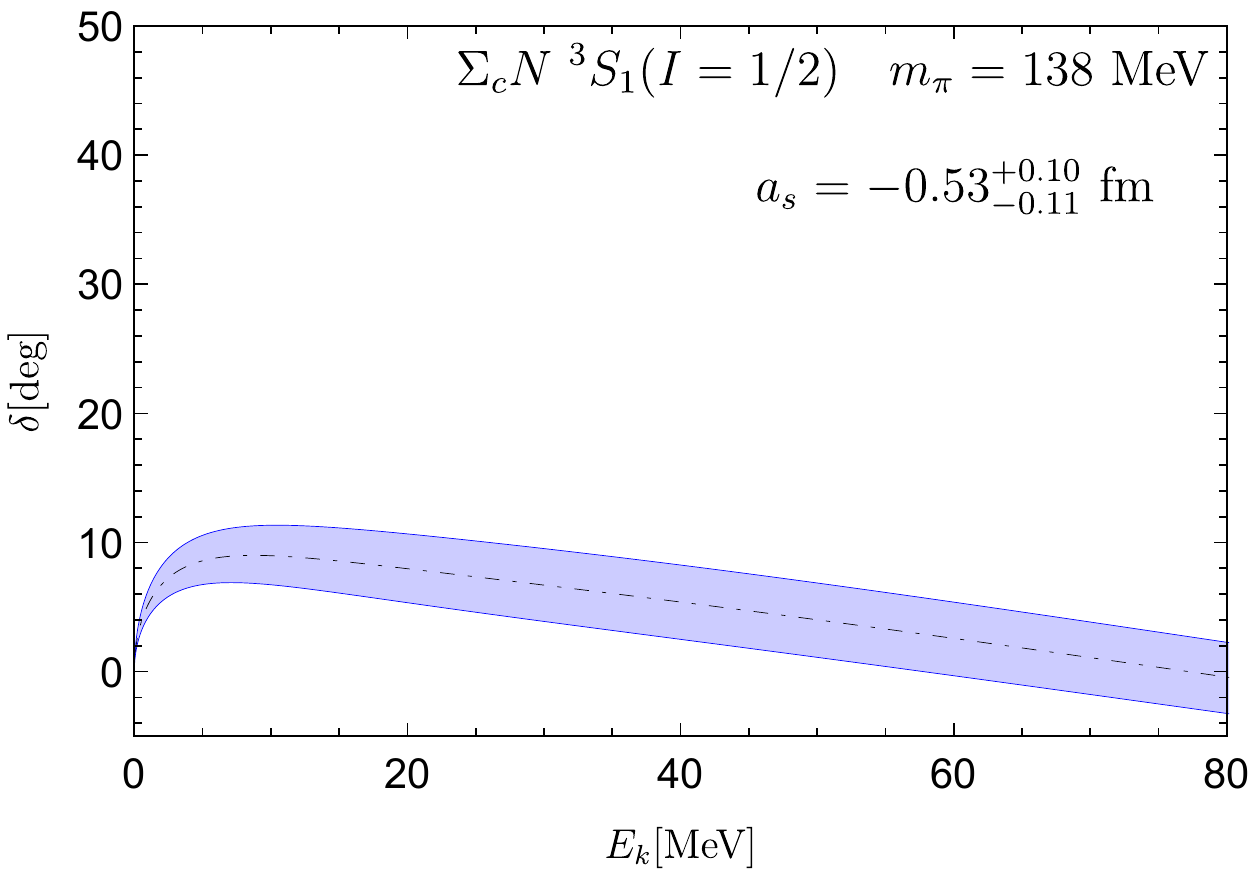}
    \caption{The chiral extrapolation of phase shift in scenario I with $n=1$ and $\Lambda=0.8$ GeV. The green and red points with error bar are the phase shift from lattice QCD~\cite{Miyamoto:2017ynx}. The black solid lines are our fitting results. The third graph is the extrapolated phase shift at the physical pion mass. The blue shadow denotes the uncertainty. }\label{fig:fm7}
\end{figure*}

\begin{figure*}[!htp]
    \centering
    \includegraphics[width=0.3\textwidth]{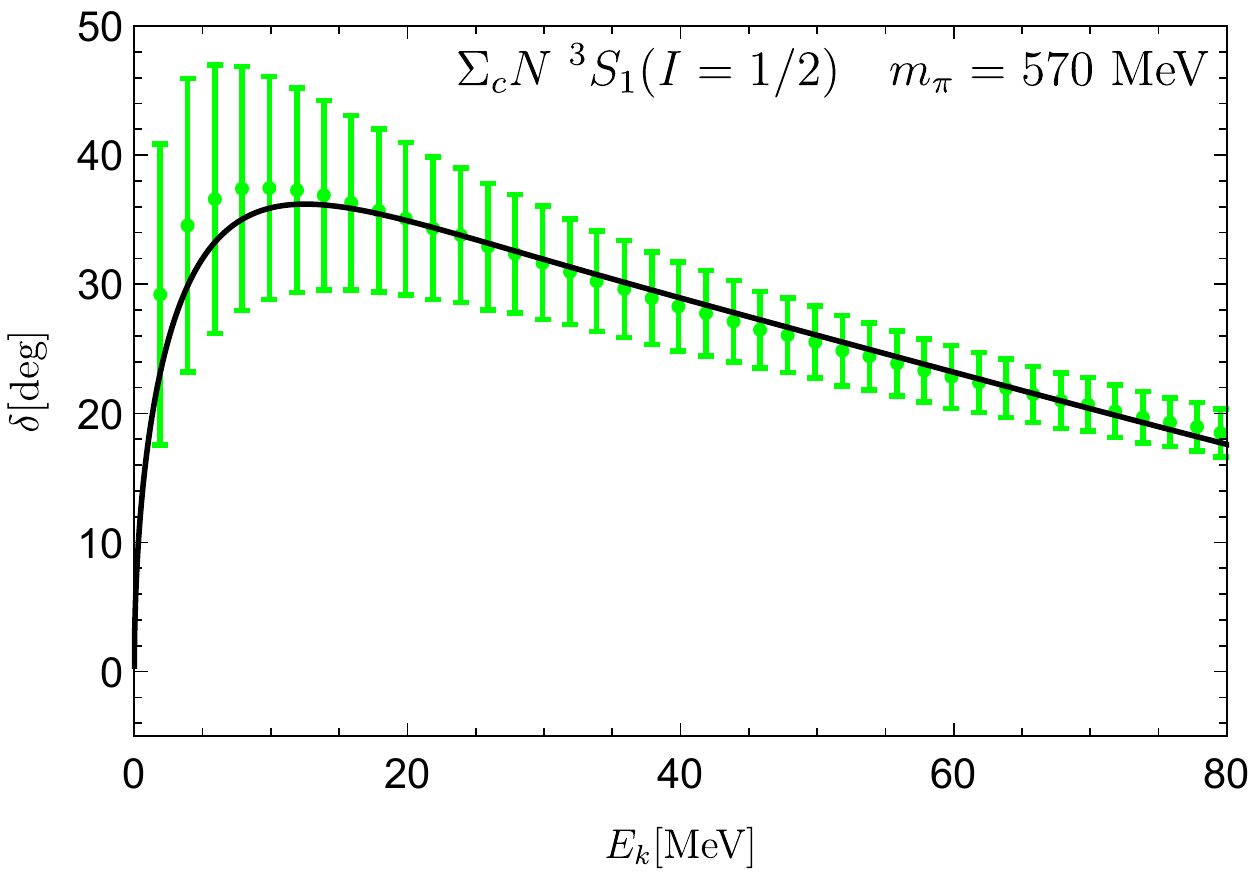}
    \includegraphics[width=0.3\textwidth]{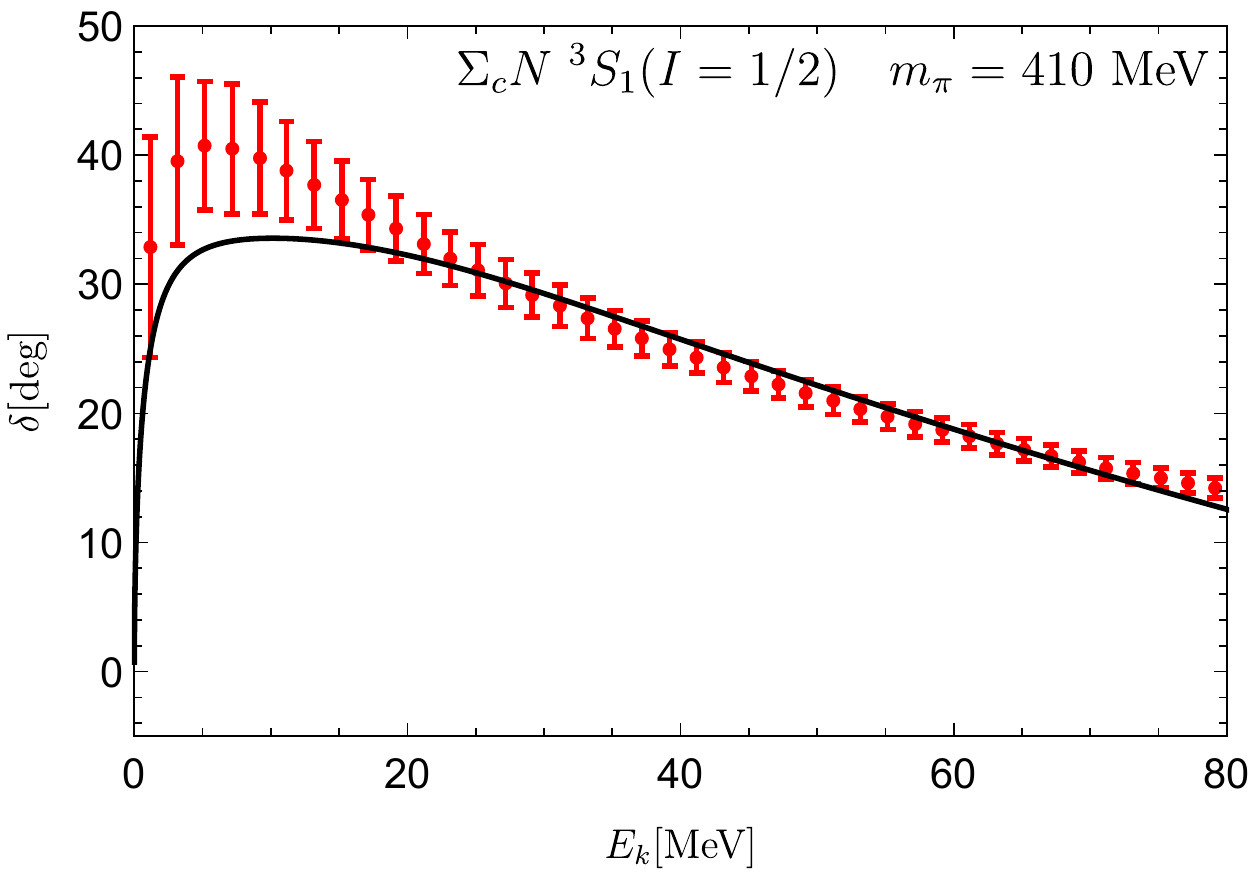}
    \includegraphics[width=0.3\textwidth]{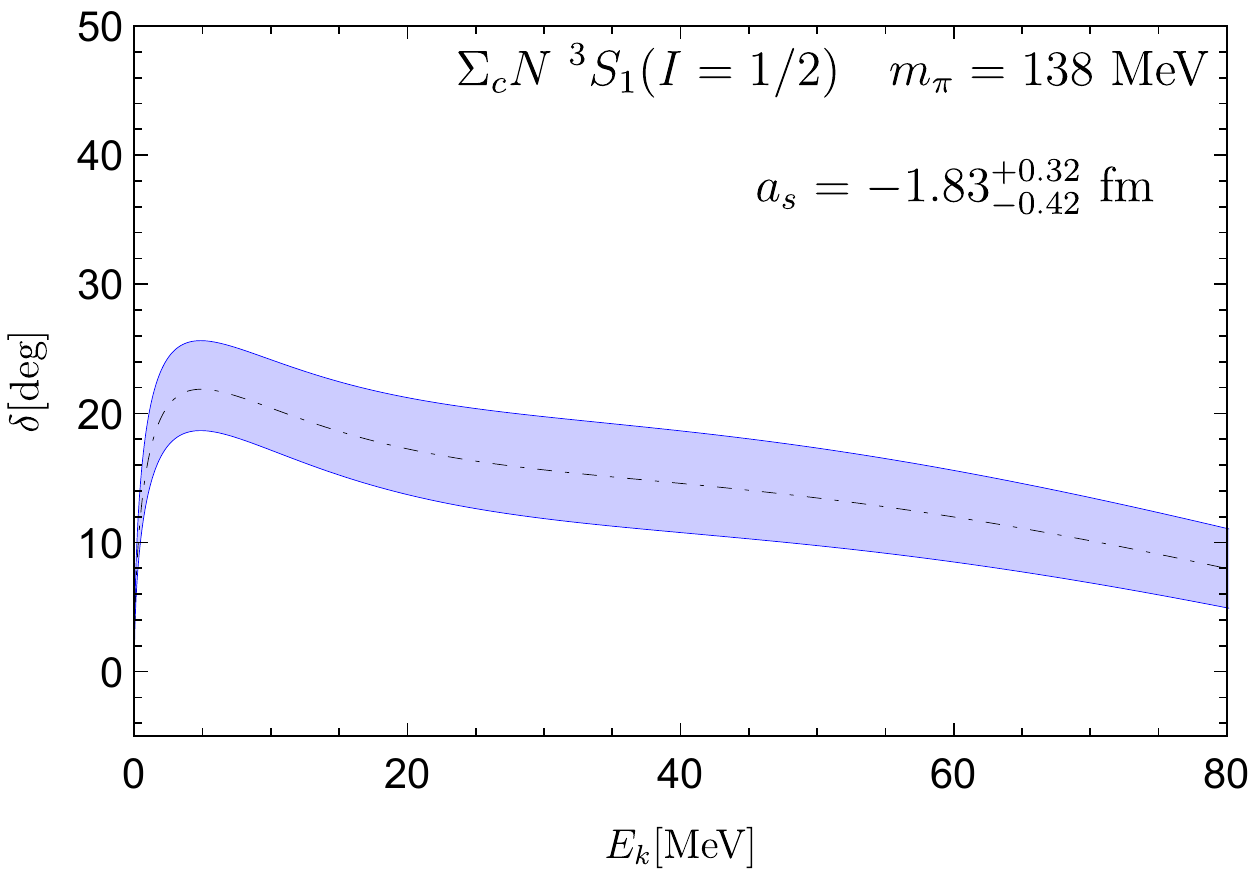}
    \caption{The chiral extrapolation of phase shift in scenario II with $n=2$ and $\Lambda=1.0$ GeV. Notations are the same as those in Fig.~\ref{fig:fm7}.}\label{fig:fm9}
\end{figure*}

\begin{figure*}[!htp]
    \centering
    \includegraphics[width=0.3\textwidth]{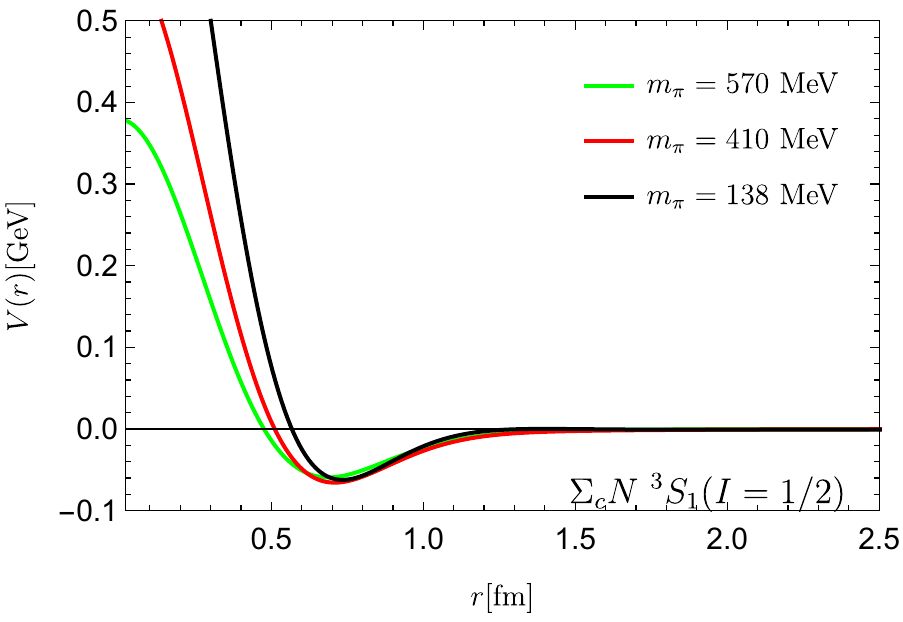}~~~~~~~~~~
    \includegraphics[width=0.3\textwidth]{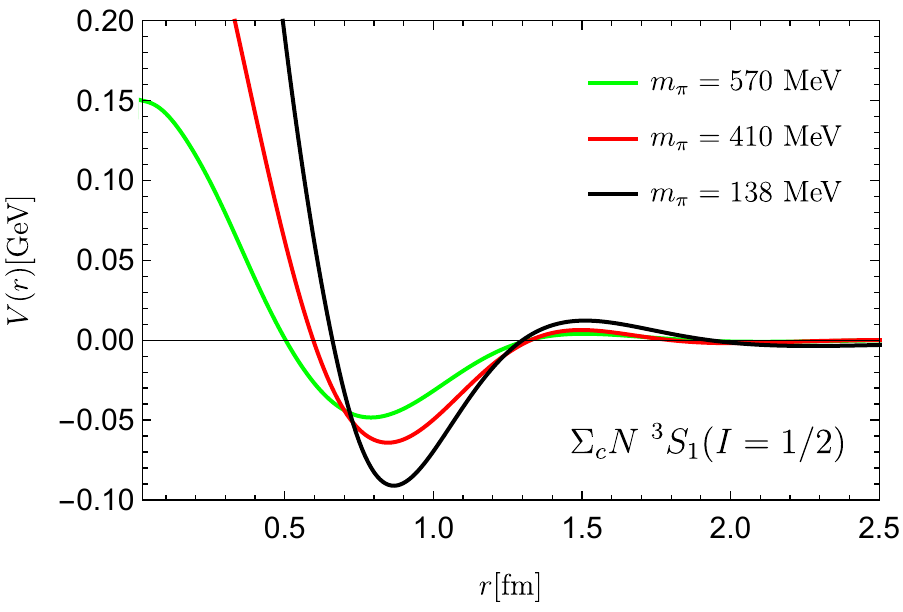}
    \caption{$\Sigma_cN$ potential in the $^3S_1(I=1/2)$ channel in  momentum space. The left one and the right one are the results in scenario I and II, respectively. }\label{fig:ptl}
\end{figure*}

The numerical results of phase shift are presented in
Figs.~\ref{fig:fm7} and \ref{fig:fm9} for scenario I and scenario
II, respectively. We fit the phase shift from lattice QCD
results~\cite{Miyamoto:2017ynx} with $m_{\pi}\approx 410$ MeV and
$m_{\pi}\approx 570$ MeV to obtain the unknown LECs $c_1$,
$\tilde{c}_1$, $c_2$ and $c_3$ [see Eq.~\eqref{eq:repara}]. Only the
$\Sigma_{c}N$ interaction with quantum number $^3S_1(I=1/2)$ is
available in lattice QCD~\cite{Miyamoto:2017ynx}. Then, we obtain
the phase shift of this channel at the physical pion mass. We show
the potential in coordinate space in Fig.~\ref{fig:ptl}.

In two scenarios, we can fit the phase shifts well with our
analytical results (see the first two graphs in Figs.~\ref{fig:fm7}
and \ref{fig:fm9}). Thus, we grasp the main features of the
$m_{\pi}$-dependent charmed baryon-nucleon interaction. We get
similar chiral extrapolation results in two scenarios with different
regulators. Potentials in two scenarios are both repulsive in the
short range and attractive in the medium range, which is the same as
the nuclear force. We also calculate the scattering lengths in two
scenarios,
\begin{eqnarray}
&a_s=-0.53^{+0.10}_{-0.11}\text{ fm     (scenario I)},\nonumber\\
&a_s=-1.83^{+0.32}_{-0.42}\text{ fm     (scenario II)}.
\end{eqnarray}
The negative scattering length indicates the attractive
$\Sigma_{c}N$ interaction in $^3S_1(I=1/2)$ channel. However, the
attraction is very weak, i.e., there do not exist the bound
solutions in both scenarios.

The different regulators in two scenarios do bring some differences
to our results. In Fig.~\ref{fig:ptl}, the short-range repulsion in
scenario I is much stronger than that in scenario II, while the
medium-range attraction in scenarios II is much stronger. Thus, both
scattering lengths in two scenarios are negative but the one in
scenario II has the larger absolute value. The differences stemming
from the choice of the regulator do not change the results
qualitatively.

In order to give the numerical results of other channels, we propose
a quark model to estimate the leading order contact interaction with
the $NN$ interaction as inputs. In HBChPT, the two-pion-exchange
diagrams mimic some heavy-meson-exchange contribution like $\rho$
and $\sigma$ in one-boson-exchange scheme. Thus, there are some
ambiguities to use specific heavy meson exchanges to calculate the
contact interaction. Meanwhile, the contact interaction obtained by
contracting the specific heavy meson exchange diagram to one single
vertex might not reproduce the $NN$ phase shift well. To avoid the
ambiguity and reduce the uncertainty, we do not consider the
specific exchanged mesons but introduce the contact interaction at
the quark level. The SU(3) flavor symmetry is used to reduce the
number of coupling constants. The unknown coupling constants are
determined by the $NN$ interaction. In this way, we could make use
of flavor symmetry to build a bridge between the $\Sigma_cN$ and
$NN$ interactions. The $\Sigma_{c}N$ potentials are shown in
Fig.~\ref{fig:qmptl}. The details about the quark model are given in
Appendix~\ref{app:qm}.

\section{Conclusion}\label{sec:conclud}

In summary, we adopt the heavy baryon chiral perturbation theory to
calculate the $S$-wave $\Sigma_{c}N$ interaction to the
next-to-leading order. We consider the leading order contact term
and one-pion-exchange interaction, the next-to-leading order contact
term and two-pion-exchange contribution. At the next-to-leading
order, we also include the renormalization of vertices, masses and
wave functions. In the calculation, the diagrams with $\Sigma_{c}^*$
as the intermediate states are taken into consideration. We use our
pion mass dependent results to fit the $\Sigma_cN~[^3S_1(I=1/2)]$
phase shifts from HAL QCD with $m_{\pi}\approx 410$ MeV and
$m_{\pi}\approx 570$ MeV, and then extrapolate the interaction to
the physical pion mass.

We choose two different regulators to give the numerical results. In
both scenarios, we can fit the phase shifts very well and obtain the
similar interaction at the physical pion mass. The potential for
$\Sigma_cN~[^3S_1(I=1/2)]$ is weakly attractive in the medium range
and repulsive in the short range. The scattering length for this
channel is negative, but the attraction is too weak to form a bound
state.

Extrapolation of lattice QCD simulation with ChPT works well when
the pion mass is as low as 300-400 \text{MeV}. When $m_{\pi}$ moves
very far away from the chiral limit, the higher order terms of the
chiral expansion would dominate the truncated expansion making the
extrapolation untenable. Fortunately, we adopt some
$m_{\pi}$-dependence directly from the lattice simulation like
$f_{\pi}$, $g_A$ $M_N$, and $M_{\Sigma_c}$, which makes our
extrapolation more reliable even for $m_{\pi}\approx 570$ MeV. There
are some approaches which can depict the pion mass dependence beyond
the perturbative chiral regime, like the cloudy bag
model~\cite{Leinweber:1998ej,Leinweber:1999ig}.

In Appendix~\ref{app:qm}, we propose a quark model to estimate the
leading order contact interaction for the $\Sigma_cN$ systems.
Rather than contracting the specific heavy meson exchange to one
point, we assume the local interaction at the quark level. We use
SU(3) flavor symmetry to reduce the coupling constants at the quark
level and adopt the $NN$ contact interaction as input. We calculate
the $\Sigma_cN$ interaction in four $S$-wave channels with the
one-pion-exchange interaction, two-pion exchange interaction and
contact interaction estimated by the quark model. The $^1S_0(I=3/2)$
channel is the most attractive one and the only one that has the
$\Sigma_cN$ bound solution.

In this work, we do not consider the couple-channel effect between
$\Sigma_cN$ and $\Lambda_cN$ channels, and the $S$-$D$ wave mixing
due to the tensor force. In the future, investigation on the $Y_cN$
interaction could be promoted by calculating the $\Lambda_cN$
scattering, including the $\Delta$ as intermediate states and
considering the $S$-$D$ wave mixing.  The framework in this work can
be extend to extrapolate the $\Sigma_c^{(*)}\bar{D}^{(*)}$
interaction to investigate the $P_c$ states in the future.

\section*{ACKNOWLEDGMENTS}

L. Meng is very grateful to G. J. Wang, X. Z. Weng, X. L. Chen and
W. Z. Deng for very helpful discussions. This project is supported
by the National Natural Science Foundation of China under Grant
11975033.

\begin{appendix}

    \section{Using quark model to estimate the contact interaction}\label{app:qm}

In Ref.~\cite{Miyamoto:2017ynx}, only the $\Sigma_{c}N$ interaction
with $^3S_1$ and $I=1/2$ was calculated. In this section, we use the
quark model to estimate the leading order contact terms and give the
numerical results for all the $S$-wave $\Sigma_cN$ interaction.

One purpose to introduce the contact terms is to include the
heavy-meson-exchange interaction after integrating out the meson
mass. At the quark level, we introduce the contact interaction as,
    \begin{eqnarray}
\mathcal{V}_{qq}=c_{S}(1+3\bm{\bm{\tau}}_{1}\cdot\bm{\bm{\tau}}_{2})+c_{T}(1+3\bm{\bm{\tau}}_{1}\cdot\bm{\bm{\tau}}_{2})\bm{\sigma}_{1}\cdot\bm{\sigma}_{2}.\quad
\label{qmlag}
    \end{eqnarray}
    We assume the local interaction is an approximation of the heavy-meson-exchange contribution at quark level when the meson masses are integrated out. The exchanged isospin singlet and isospin triplet belong to the same multiplets in the SU(3) flavor symmetry. Thus, they share the same coupling constants. Since we focus on the interaction of $u/d$ quarks, we can write the exchanged mesons as
    \begin{eqnarray}
    \mathbb{M}=\bm{a\cdot\tau}+\frac{1}{\sqrt{3}}f,
    \end{eqnarray}
    where $a$ and $f$ generally represent the isospin triplet and isospin singlet, respectively. Thus, there are only two independent coupling constants in Eq.~\eqref{qmlag}.

With the quark level interaction, we can calculate the $NN$ and
$\Sigma_{c}N$ contact interaction. We present the relevant matrix
elements in Table~\ref{tab:me}. Thus, we could relate the unknown
LECs for $\Sigma_cN$ in Eq.~\eqref{lag:loctc} to the $NN$ contact
interaction with the quark model as a bridge.

For the $NN$ system, the leading order contact interaction
reads~\cite{Machleidt:2011zz},
    \begin{eqnarray}
    {\cal V}_{NN}^{(0)}=C_{S}+C_{T}\bm{\sigma_{1}\cdot\sigma_{2}},\label{lag:nn}
    \end{eqnarray}
where $C_S$ and $C_T$ are the LECs for central potential and
spin-spin interaction, respectively. In principle, one also needs to
construct the isospin-isospin interaction. However, the two nucleons
are identical particles and satisfy the Pauli principle. For the
$S$-wave two nucleon system, once the total spin is fixed, the total
isospin could be determined consequently. Thus, the isospin-isospin
interaction could be absorbed into $C_S$ and $C_T$. The value of
$C_S$ and $C_T$ have been determined by fitting the nucleon
scattering phase shift, which is less dependent on the cutoff in the
regulator. We take $C_S=-100 \text{ GeV}^{-2}$ and $C_T=6.5 \text{
GeV}^{-2}$~\cite{Machleidt:2011zz}. The LECs in the leading order
$\Sigma_cN$ contact Lagrangians [see Eq.~\eqref{lag:loctc}] are then
determined as
    \begin{eqnarray}
    &C_3=-5.21\text{ GeV}^{-2},\quad C_4=-5.21\text{ GeV}^{-2},\nonumber\\
    &\tilde{C}_{3}=5.19\text{ GeV}^{-2},\quad \tilde{C}_{4}=25.97\text{ GeV}^{-2}.
    \end{eqnarray}

\begin{figure*}[!htp]
    \centering
    \includegraphics[width=0.3\textwidth]{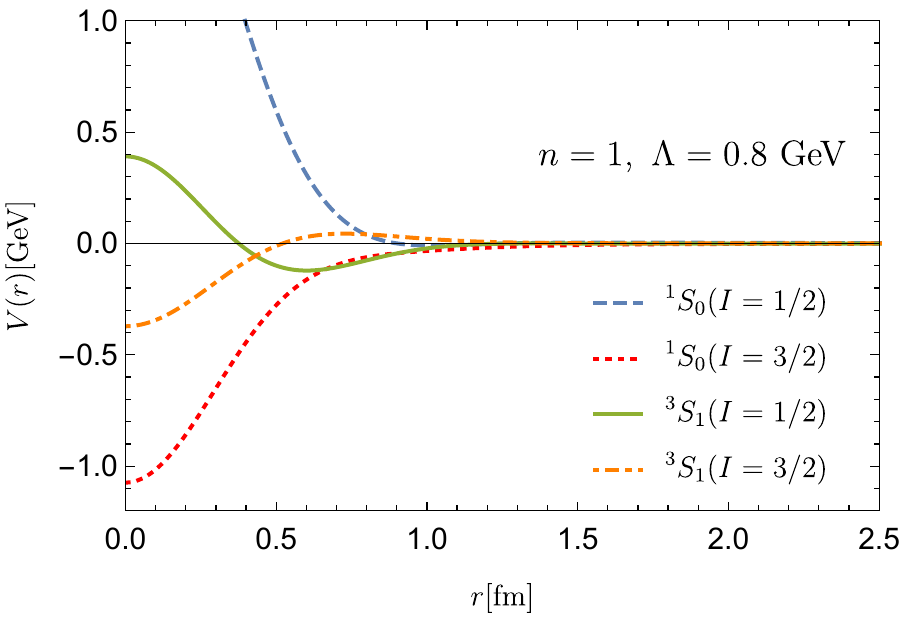}~~~~~~~~~~
    \includegraphics[width=0.3\textwidth]{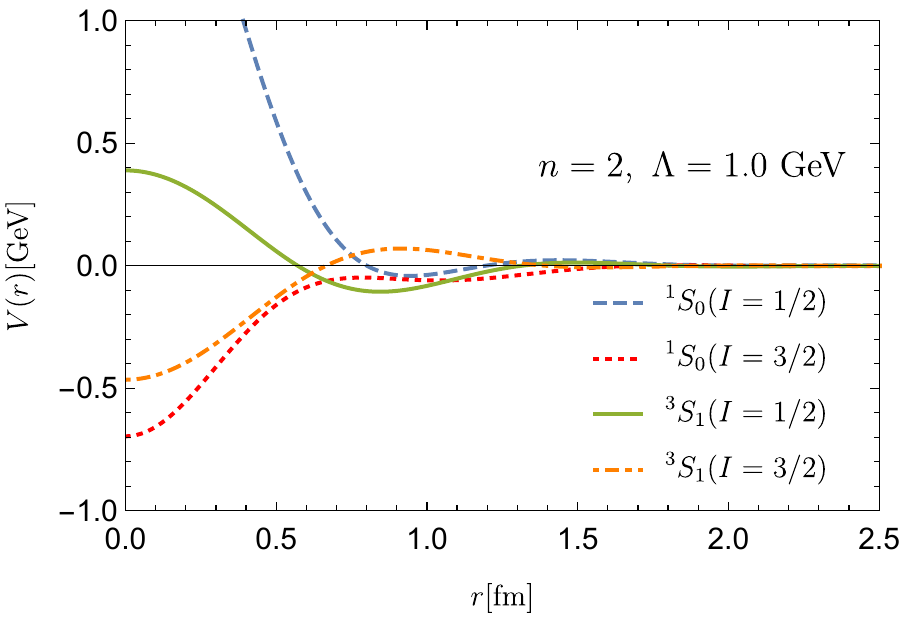}
    \caption{$\Sigma_cN $ potentials obtained by combining the chiral effective field theory and quark model. The potentials include the leading order contact term estimated in quark model with the $NN$ contact interaction as inputs, the one-pion-exchange interaction and the two-pion-exchange contribution.}\label{fig:qmptl}
\end{figure*}

The $\Sigma_{c}N$ potentials in coordinate space are presented in
Fig.~\ref{fig:qmptl}. The contribution of the leading order contact
term, the one-pion-exchange interaction and the two-pion-exchange
contribution are included. With the quark model as a bridge, we
could determine the $\Sigma_cN$ effective potentials for all four
$S$-wave channels. The scenarios with different regulators give the
similar results.

For the $^3S_1 (I=1/2)$ channel of $\Sigma_cN$, the medium-range
interaction is attractive and the short-range interaction is
repulsive, which is in accordance with the chiral extrapolation of
lattice QCD results. There does not exist the bound solution in this
channel. The  $^1S_0 (I=1/2)$ channel is strongly repulsive. For the
$I=3/2$ channel, the $\Sigma_cN$ interaction with quantum number
$^3S_1$ is repulsive in the medium range but attractive in the short
range. There is no bound solution in this channel. The $^1S_0
(I=3/2)$ channel is the most attractive one. We find a bound
solution in this channel,
\begin{eqnarray}
&E=-172\text{ MeV (Scenario I)}, \nonumber\\
&E=-48\text{ MeV (Scenario II)}.
\end{eqnarray}

\begin{table}
    \caption{The matrix elements of the  operator $\sum_{i\in h_a,j\in h_b}O_{ij} $, where $h_a$ and $h_b$ are two hadrons. $O_{ij}$ is the two-body interaction operator between quarks.}\label{tab:me}
    \setlength{\tabcolsep}{3.2mm}
        \begin{tabular}{ccccc}
    \toprule[0.5pt]\toprule[0.5pt]
            $O_{ij}$ & $1_{ij}$ & $\bm{\sigma}_{i}\cdot\bm{\sigma}_{j}$ & $\bm{\tau}_{i}\cdot\bm{\tau}_{j}$ & $\bm{\sigma}_{i}\cdot\bm{\sigma}_{j}\bm{\tau}_{i}\cdot\bm{\tau}_{j}$\tabularnewline
    \midrule[0.5pt]
            $[NN]_{S=0}^{I=1}$ & $9$ & $-3$ & $1$ & $-\frac{25}{3}$\tabularnewline
                    $[NN]_{S=1}^{I=0}$ & $9$ & $1$ & $-3$ & $-\frac{25}{3}$\tabularnewline
            $[\Sigma_{c}N]_{S=1}^{I=\frac{3}{2}}$ & $6$ & $\frac{4}{3}$ & $2$ & $\frac{20}{9}$\tabularnewline
            $[\Sigma_{c}N]_{S=0}^{I=\frac{3}{2}}$ & $6$ & $-4$ & $2$ & $-\frac{20}{3}$\tabularnewline
                $[\Sigma_{c}N]_{S=1}^{I=\frac{1}{2}}$ & $6$ & $\frac{4}{3}$ & $-4$ & $-\frac{40}{9}$\tabularnewline
            $[\Sigma_{c}N]_{S=0}^{I=\frac{1}{2}}$ & $6$ & $-4$ & $-4$ & $\frac{40}{3}$\tabularnewline
        \bottomrule[0.5pt]\bottomrule[0.5pt]
        \end{tabular}
    \end{table}
The behaviors of the effective potentials for
$S$-wave $\Sigma_cN$ channels are in accordance with the refined
quark model calculation~\cite{Garcilazo:2019ryw}. The approach
applied to the contact interaction could also be used to investigate
the $P_c$ states~\cite{Meng:2019ilv,Wang:2019ato}, in which the
$\Lambda_c\bar{D}^{(*)}$ interaction can be related to the
$\Sigma_c^{(*)}\bar{D}^{(*)}$ ones.

\section{Definitions of the loop integrals}\label{app:super}
We will use the ``M$x$B$y$'' to denote the integrals with $x$ light
meson propagators and $y$ heavy baryon propagators in the following.
\begin{widetext}
    \begin{itemize}
        \item M1B0
        \begin{equation}
        i\int\frac{d^{d}l\lambda^{4-d}}{(2\pi)^{d}}\frac{\{1,l^{\alpha},l^{\alpha}l^{\beta}\}}{l^{2}-m^{2}+i\varepsilon}\equiv\left\{
        J_{0}^{c},0,g^{\alpha\beta}J_{21}^{c}\right\} (m),
        \end{equation}
        \item M2B0
        \begin{equation}
        i\int\frac{d^{d}l\lambda^{4-d}}{(2\pi)^{d}}\frac{\{1,l^{\alpha},l^{\alpha}l^{\beta},l^{\alpha}l^{\beta}l^{\gamma}\}}{\left(l^{2}-m^{2}+i\varepsilon\right)\left[(l+q)^{2}-m^{2}+i\varepsilon\right]}\equiv\left\{
        J_{0}^{F},q^{\alpha}J_{11}^{F},q^{\alpha}q^{\beta}J_{21}^{F}+g^{\alpha\beta}J_{22}^{F},(g\vee
        q)J_{31}^{F}+q^{\alpha}q^{\beta}q^{\gamma}J_{32}^{F}\right\} (m,q),
        \end{equation}
        \item M1B1
        \begin{equation}
        i\int\frac{d^{d}l\lambda^{4-d}}{(2\pi)^{d}}\frac{\{1,l^{\alpha},l^{\alpha}l^{\beta},l^{\alpha}l^{\beta}l^{\gamma}\}}{\left(v\cdot
            l+\omega+i\varepsilon\right)\left(l^{2}-m^{2}+i\varepsilon\right)}\equiv\left\{
        J_{0}^{a},v^{\alpha}J_{11}^{a},v^{\alpha}v^{\beta}J_{21}^{a}+g^{\alpha\beta}J_{22}^{a},(g\vee
        v)J_{31}^{a}+v^{\alpha}v^{\beta}v^{\gamma}J_{32}^{a}\right\}
        (m,\omega),
        \end{equation}
        \item M2B1
        \begin{eqnarray}
        &&i\int\frac{d^{d}l\lambda^{4-d}}{(2\pi)^{d}}\frac{\{1,l^{\alpha},l^{\alpha}l^{\beta},l^{\alpha}l^{\beta}l^{\gamma},l^{\alpha}l^{\beta}l^{\gamma}l^{\delta}\}}{\left(v\cdot l+\omega+i\varepsilon\right)\left(l^{2}-m^{2}+i\varepsilon\right)\left[(l+q)^{2}-m^{2}+i\varepsilon\right]}\equiv\Big\{ J_{0}^{T},q^{\alpha}J_{11}^{T}+v^{\alpha}J_{12}^{T},g^{\alpha\beta}J_{21}^{T}+q^{\alpha}q^{\beta}J_{22}^{T}+v^{\alpha}v^{\beta}J_{23}^{T}\nonumber \\
        &&  \quad+(q\vee v)J_{24}^{T},(g\vee q)J_{31}^{T}+q^{\alpha}q^{\beta}q^{\gamma}J_{32}^{T}+(q^{2}\vee v)J_{33}^{T}+(g\vee v)J_{34}^{T}+(q\vee v^{2})J_{35}^{T}+v^{\alpha}v^{\beta}v^{\gamma}J_{36}^{T},(g\vee g)J_{41}^{T} \nonumber \\
        &&\quad+(g\vee q^{2})J_{42}^{T}+q^{\alpha}q^{\beta}q^{\gamma}q^{\delta}J_{43}^{T}+(g\vee v^{2})J_{44}^{T}+v^{\alpha}v^{\beta}v^{\gamma}v^{\delta}J_{45}^{T}+(q^{3}\vee v)J_{46}^{T}+(q^{2}\vee v^{2})J_{47}^{T}+(q\vee v^{3})J_{48}^{T} \nonumber \\
        &&  \quad+(g\vee q\vee v)J_{49}^{T}\Big\}(m,\omega,q),
        \end{eqnarray}
        \item M1B2
        \begin{eqnarray}
        &&i\int\frac{d^{D}l\lambda^{4-D}}{(2\pi)^{D}}\frac{\{1,l^{\alpha},l^{\alpha}l^{\beta},l^{\alpha}l^{\beta}l^{\gamma}\}}{\left(v\cdot l+\omega_{1}+i\varepsilon\right)\left[(+/-)v\cdot l+\omega_{2}+i\varepsilon\right]\left(l^{2}-m^{2}+i\varepsilon\right)} \nonumber\\
        &&\equiv\Big\{ J_{0}^{g/h},v^{\alpha}J_{11}^{g/h},v^{\alpha}v^{\beta}J_{21}^{g/h}+g^{\alpha\beta}J_{22}^{g/h},
        (g\vee v)J_{31}^{g/h}+v^{\alpha}v^{\beta}v^{\gamma}J_{32}^{g/h}\Big\} (m,\omega_{1},\omega_{2}),
        \end{eqnarray}
        \item M2B2
        \begin{eqnarray}
        &&i\int\frac{d^{d}l\lambda^{4-d}}{(2\pi)^{d}}\frac{\{1,l^{\alpha},l^{\alpha}l^{\beta},l^{\alpha}l^{\beta}l^{\gamma},l^{\alpha}l^{\beta}l^{\gamma}l^{\delta}\}}{\left(v\cdot l+\omega_{1}+i\varepsilon\right)\left[(+/-)v\cdot l+\omega_{2}+i\varepsilon\right]\left(l^{2}-m^{2}+i\varepsilon\right)\left[(l+q)^{2}-m^{2}+i\varepsilon\right]}\equiv\Big\{ J_{0}^{R/B},q^{\alpha}J_{11}^{R/B}+v^{\alpha}J_{12}^{R/B},\nonumber\\
        &&\quad g^{\alpha\beta}J_{21}^{R/B} +q^{\alpha}q^{\beta}J_{22}^{R/B}+v^{\alpha}v^{\beta}J_{23}^{R/B}+(q\vee v)J_{24}^{R/B},(g\vee q)J_{31}^{R/B}+q^{\alpha}q^{\beta}q^{\gamma}J_{32}^{R/B}+(q^{2}\vee v)J_{33}^{R/B}+(g\vee v)J_{34}^{R/B}\nonumber\\
        &&  \quad+(q\vee v^{2})J_{35}^{R/B}+v^{\alpha}v^{\beta}v^{\gamma}J_{36}^{R/B},(g\vee g)J_{41}^{R/B}+(g\vee q^{2})J_{42}^{R/B}+q^{\alpha}q^{\beta}q^{\gamma}q^{\delta}J_{43}^{R/B}+(g\vee v^{2})J_{44}^{R/B}+v^{\alpha}v^{\beta}v^{\gamma}v^{\delta}J_{45}^{R/B}\nonumber\\
        &&\quad +(q^{3}\vee v)J_{46}^{R/B}+(q^{2}\vee v^{2})J_{47}^{R/B}+(q\vee v^{3})J_{48}^{R/B}+(g\vee q\vee v)J_{49}^{R/B}\Big\}(m,\omega_{1},\omega_{2},q),
        \end{eqnarray}
    \end{itemize}
    We use the representation $X\vee Y \vee Z...$ to denote the symmetrized
    tensor structure for simplicity. For example,
    \begin{eqnarray}
    q\vee v &\equiv&
    q^{\alpha}v^{\beta}+q^{\beta}v^{\alpha},\nonumber\\
    g\vee q &\equiv&
    g^{\alpha\beta}q^{\gamma}+g^{\alpha\gamma}q^{\beta}+g^{\gamma\beta}q^{\alpha},\nonumber\\
    g\vee g&\equiv& g^{\alpha\beta}g^{\gamma
        \delta}+g^{\alpha\gamma}g^{\beta\delta}+g^{\alpha\delta}g^{\beta\gamma}.
    \end{eqnarray}
    The final results of these integrals can be found in Refs.~\cite{Meng:2019ilv,Wang:2019ato}.
\end{widetext}

\end{appendix}

\vfil \thispagestyle{empty}

\newpage
\bibliography{ref}

\end{document}